\documentclass[showpacs]{revtex4}
\usepackage{graphicx}
\usepackage{dcolumn}
\usepackage{amsmath}

\begin{document}

%%%%%%%%%%%%%%%%%%%%%%%%%%%%%%%%%%%%%%%%%%%%%%%%%%%%%%%%%%%%%%%%%%%%%%%%%%

\title{ Noncommutative Quantum Cosmology}

%%%%%%%%%%%%%%%%%%%%%%%%%%%%%%%%%%%%%%%%%%%%%%%%%%%%%%%%%%%%%%%%%%%%%%%%%%

\author{Luis O. Pimentel\dag \, and
 C\'esar Mora\ddag \\
\dag Departamento de F\'{\i}sica,\\
Universidad Aut\'onoma Metropolitana,\\
Apartado Postal 55-534,CP 09340 M\'exico D.F., MEXICO.\\
\ddag  Departamento de Matem\'aticas,\\
UPIBI-Instituto Polit\'ecnico Nacional,\\
Av. Acueducto s/n Col. Barrio La Laguna Ticom\'an, \\
CP 07340 M\'exico DF, MEXICO.}

\date{\today}

%%%%%%%%%%%%%%%%%%%%%%%%%%%%%%%%%%%%%%%%%%%%%%%%%%%%%%%%%%%%%%%%%%%%%%%%%%
\begin{abstract}
We consider noncommutative quantum cosmology in the case of the
low-energy string effective theo\-ry. Exacts solutions are found
and compared with the commutative case.The Noncommutative quantum
cosmology is considered in the case of the low-energy string
effective theory. Exacts solutions are found and compared with the
commutative case.

\end{abstract}
%%%%%%%%%%%%%%%%%%%%%%%%%%%%%%%%%%%%%%%%%%%%%%%%%%%%%%%%%%%%%%%%%%%%%%%%%%

\maketitle \vskip -.5truecm

%\maketitle

%%%%%%%%%%%%%%%%%%%%%%%%%%%%%%%%%%%%%%%%%%%%%%%%%%%%%%%%%%%%%%%%%%%%%%%

\section{Introduction}

There is a renewed interest on non commutative theories in
physics, one of the reasons for that is that noncommutativity
between coordinates appears in string theory: in the toroidal
compactification of Matrix Theory  \cite{connes} and in open
string theory with a B-field background \cite{openB}.
Noncommutative classical \cite{cnc} and quantum  mechanics
\cite{qnc} has been considered. The effects of noncommutativity
has been studied  also in gravity \cite{gnc}. In the field of
quantum cosmology, noncommutativity of the minisuperspace
coordinates  was introduce by Compean {\it et al.}\cite{compean}.
They studied the Kantowski-Sachs metric and were able to find the
exact wave function, with which wave packets were constructed and
found new quantum states that ``compete'' to be the most probable
state, in clear contrast with the commutative case. A tunnelling
process could be possible among these different states for the
Universe. We want to explore the effects of noncommutativity in
quantum cosmology when matter fields are included.

%%%%%%%%%%%%%%%%%%%%%%%%%%%%%%%%%%%%%%%%%%%%%%%%%%%%%%%%%%%%%%%%%%%%%%%

\section{The low-energy string effective quantum cosmology}

%%%%%%%%%%%%%%%%%%%%%%%%%%%%%%%%%%%%%%%%%%%%%%%%%%%%%%%%%%%%%%%%%%%%%%%

In this work we want to study noncommutative cosmology with some
matter included and not just pure gravity as was the case of
\cite{compean}. In order to have an exact solution we must
consider a geometry with high symmetry, the
Robertson-Walker-Friedmann models.For the matter content of our
model we will consider the low energy limit of string theory and
we will have an scalar field besides the gravitational degrees of
freedom. Some other gravitational theories of the scalar-tensor
type are under consideration \cite{pm}. At low energy, the
tree-level, $(3+1)$-dimensional string effective action can be
written as \cite{Gasperini}
\begin{equation}
S=-\frac{1}{2\lambda_s}\int
d^4x\sqrt{-g}e^{-\phi}(R+\partial_\mu\phi\partial^\mu\phi+V).
\end{equation}
Here $\phi$ is the dilaton field,
$a(t)=\exp\big[{\beta(t)/\sqrt{3}}\big]$, $\lambda_s$ is the
fundamental string length parameter governing the high-derivative
expansion of the action and $V$ is a possible dilaton potential.
When we consider this theory in the metric of isotropic and
homogeneous spacetime, after integrating by parts, and using the
convenient time parametrization $dt= e^{-\bar{\phi}}d\tau$,
reduces to (in the gauge $g_{00}=1$)
\begin{equation}
S=-\frac{\lambda_s}{2}\int d\tau\Big(\bar{\phi}^{\prime 2}
-\bar{\beta}^{\prime 2} + Ve^{-2\bar{\phi}}\Big),
\end{equation}
where
\begin{equation}
\bar{\phi}=\phi-\ln\int(d^3x/\lambda^3_s)-\sqrt{3}\beta.
\end{equation}
The Hamiltonian of the system is
\begin{equation}
H=\frac{1}{2\lambda_s}\left(
\Pi_\beta^2-\Pi_\phi^2+\lambda_s^2Ve^{2\bar{\phi}}\right),
\end{equation}
where the canonical conjugate momenta are,
\begin{equation}
\Pi_\beta=\lambda_s\beta^\prime, \qquad
\Pi_\phi=-\lambda_s\bar{\phi}^\prime.
\end{equation}
The corresponding Wheeler-DeWitt equation, in a particular factor
ordering is
\begin{equation}
\frac{1}{2\lambda_s}\left[\frac{\partial^2}{\partial\bar{\phi}^2}-
\frac{\partial^2}{\partial\beta^2}
+\lambda_s^2V(\bar{\phi},\beta)e^{-2\bar{\phi}}\right]\psi(\phi,\beta)=0.
\label{wdw-1}
\end{equation}
We shall assume $V=V(\phi)$ in order to separate variables. We
consider two simple cases of the potential as toy models that allow us to obtain exact
solutions\\

\noindent 1) Case: $V=-V_0e^{4\bar{\phi}}$\\

\noindent Therefore the solution of the WDW equation (\ref{wdw-1})
is
\begin{equation}
\psi_\nu(\bar\phi,\beta)=Ce^{i\nu\beta}Y_{i\nu}\left(\lambda_s\sqrt{V_0}e^{\bar\phi}\right),
\end{equation}
where $Y_{i\nu}$ is the second class Bessel function.\\

\noindent 2) Case: $V=-V_0$\\

\noindent Now the wave function is
\begin{equation}
\psi_\nu(\bar\phi,\beta)=Ce^{i\nu\beta}K_{i\nu}\left(\lambda_s\sqrt{V_0}e^{-2\bar\phi}\right),
\end{equation}
where $K_{i\nu}$ is the modified  Bessel function. We can
construct wormhole type solutions by means integrating over the
separation constant $\nu$,
\begin{eqnarray}
\psi_{WH}(\bar\phi,\beta)&=&\int_{-\infty}^{+\infty}
e^{i\nu(\beta+\mu)}K_{i\nu}\left(\lambda_s\sqrt{V_0}e^{\bar\phi}\right)d\nu,
\nonumber\\
&=&e^{-\lambda_s\sqrt{V_0}e^{\bar\phi}\cosh[2\beta+\mu]}.
\end{eqnarray}
where $\mu=const$. For the noncommutative quantum cosmology model,
we will assume the the "cartesian coordinates" $\bar\phi$ and
$\beta$ of the Robertson-Walker minisuperspace obey a kind of
commutation relation \cite{Gamboa},
\begin{equation}
[\bar\phi,\beta]=i\theta.
\end{equation}
This is a particular ansatz in these configuration coordinates.
The deformation of minisuperspace can be studied in terms of Moyal
\cite{Moyal} product,
\begin{equation}
f(\bar\phi,\beta)\star g(\bar\phi,\beta)= f(\bar\phi,\beta)
\exp{\left[i\frac{\theta}{2}\left(\overleftarrow{\frac{\partial}{\partial\bar\phi}}
\overrightarrow{\frac{\partial}{\partial\beta}}-\overleftarrow{\frac{\partial}{\partial\beta}}
\overrightarrow{\frac{\partial}{\partial\bar\phi}}\right)\right]}g(\bar\phi,\beta).
\end{equation}
Then the noncommutative WDW equation is
\begin{equation}
\frac{1}{2\lambda_s}\star\left[\frac{\partial^2}{\partial\bar{\phi}^2}-
\frac{\partial^2}{\partial\beta^2}+\lambda_s^2V(\bar{\phi},\beta)e^{-2\bar{\phi}}\right]
\star\psi(\phi,\beta)=0.
\label{wdw-2}
\end{equation}
It is possible to reformulate this equation in terms of
commutative variables and the ordinary product of functions, if
new variables are introduced $\bar\phi\to
\bar\phi-\frac{1}{2}\theta\Pi_\beta$ and $\beta\to
\beta+\frac{1}{2}\theta\Pi_{\bar\phi}$. Therefore,  the original
equation changes, with a potential modified due to these new
variables.
\begin{equation}
V(\bar\phi,\beta)\star\psi(\bar\phi,\beta)=V\left(\bar\phi-\frac{1}{2}\theta\Pi_\beta,
\beta+\frac{1}{2}\theta\Pi_{\bar\phi}\right)\psi(\bar\phi,\beta).
\end{equation}
Then, we get
\begin{equation}
\left[\frac{\partial^2}{\partial \bar\phi^2} -
\frac{\partial^2}{\partial \beta^2} + \lambda_s^2Ve^{-2\bar\phi+\theta\Pi_{\beta}}
\right]\psi(\bar\phi,\beta)=0. \label{wdw-3}
\end{equation}
In order to separate variables we propose the ansatz
\begin{equation}
\psi(\bar\phi,\beta)=e^{\nu\beta}B(\bar\phi),
\end{equation}
the operator $\Pi_\beta$ in the exponential in equation
(\ref{wdw-3}) will shift the wave function by a factor
\begin{equation}
\psi(\bar\phi,\beta-i\nu\theta)=e^{-2i\nu\theta}\psi(\bar\phi,\beta),
\end{equation}
thus $B(\bar\phi)$ must satisfy the equation
\begin{equation}
\left[\frac{d^2}{d\bar\phi^2}
+\nu^2+\lambda_s^2Ve^{-2\bar\phi-2i\nu\theta}\right]F(\bar\phi)=0.
\end{equation}
1)  $V=V_0e^{4\bar\phi}$\\

\noindent The wave function is given by
\begin{equation}
\psi(\bar\phi,\beta)=e^{i\nu\beta}Y_{i\nu}\left(\lambda_s
\sqrt{V_0}e^{\bar\phi-i\nu\theta}\right),
\end{equation}
we can see that noncommutativity induces a difference of the
arguments of the Bessel functions in the wave function.\\

\noindent 2)  $V=V_0$\\

\noindent Now the wave function is
\begin{equation}
\psi(\bar\phi,\beta)=e^{i\nu\beta}K_{i\nu}\left(\lambda_s
\sqrt{V_0}e^{-2\bar\phi-i\nu\theta}\right).
\end{equation}
Now, in order to see a direct influence of parameter $\theta$ we
construct the wave packet
\begin{equation}
\psi(\bar\phi,\beta)=\int_{-\infty}^\infty
e^{-a(\nu-b)^2}e^{i\nu\beta}K_{i\nu}\left(\lambda_s
\sqrt{V_0}e^{-2\bar\phi-i\nu\theta}\right)\,d\nu.
\end{equation}

In figures 1 and 2 we have plotted the square of the absolute
value of the wave packets for different values of the
noncommutativity parameter $\theta$ and the chosen values of $a=2$
and $b=3$. From the figure it is clear how important could be the
existence of noncommutativity at the early stages od the Universe.
At present times the experiment related to noncommutative quantum
mechanics set a very small value for the parameter $\theta$,
however those limits may not be applicable in quantum cosmology
with the high energy regime.

%%%%%%%%%%%%%%%%%%%%%%%%%%%%%%%%%%%%%%%%%%%%%%%%%%%%%%%%%%%%%%%%%%%%%%%
\begin{figure}
\begin{center}
\includegraphics[width=8 cm]{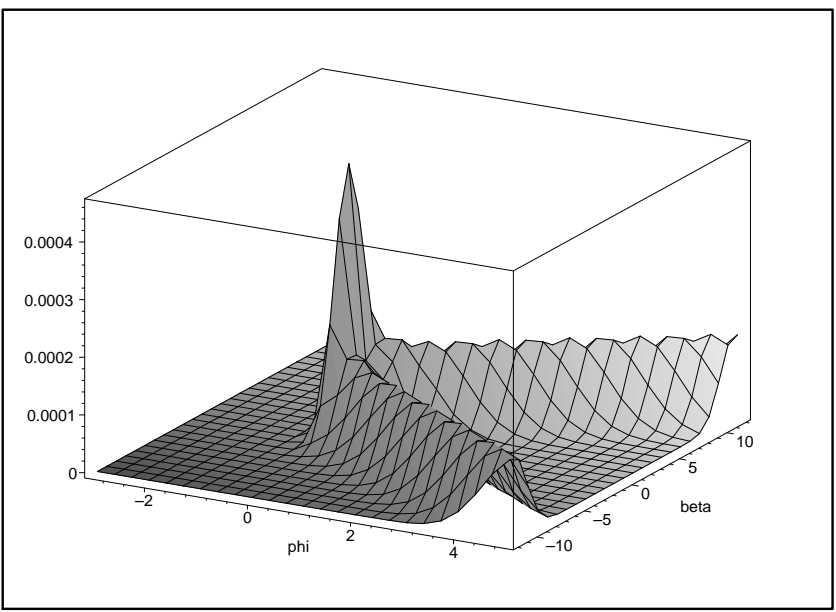}\\
\end{center}
\caption{ $|\Psi(\phi, \beta)|^2$ for $\theta=0$.} \label{wavef1}
\end{figure}
%%%%%%%%%%%%%%%%%%%%%%%%%%%%%%%%%%%%%%%%%%%%%%%%%%%%%%%%%%%%%%%%%%%%%%%

%%%%%%%%%%%%%%%%%%%%%%%%%%%%%%%%%%%%%%%%%%%%%%%%%%%%%%%%%%%%%%%%%%%%%%%
\begin{figure}
\begin{center}
\includegraphics[width=8 cm]{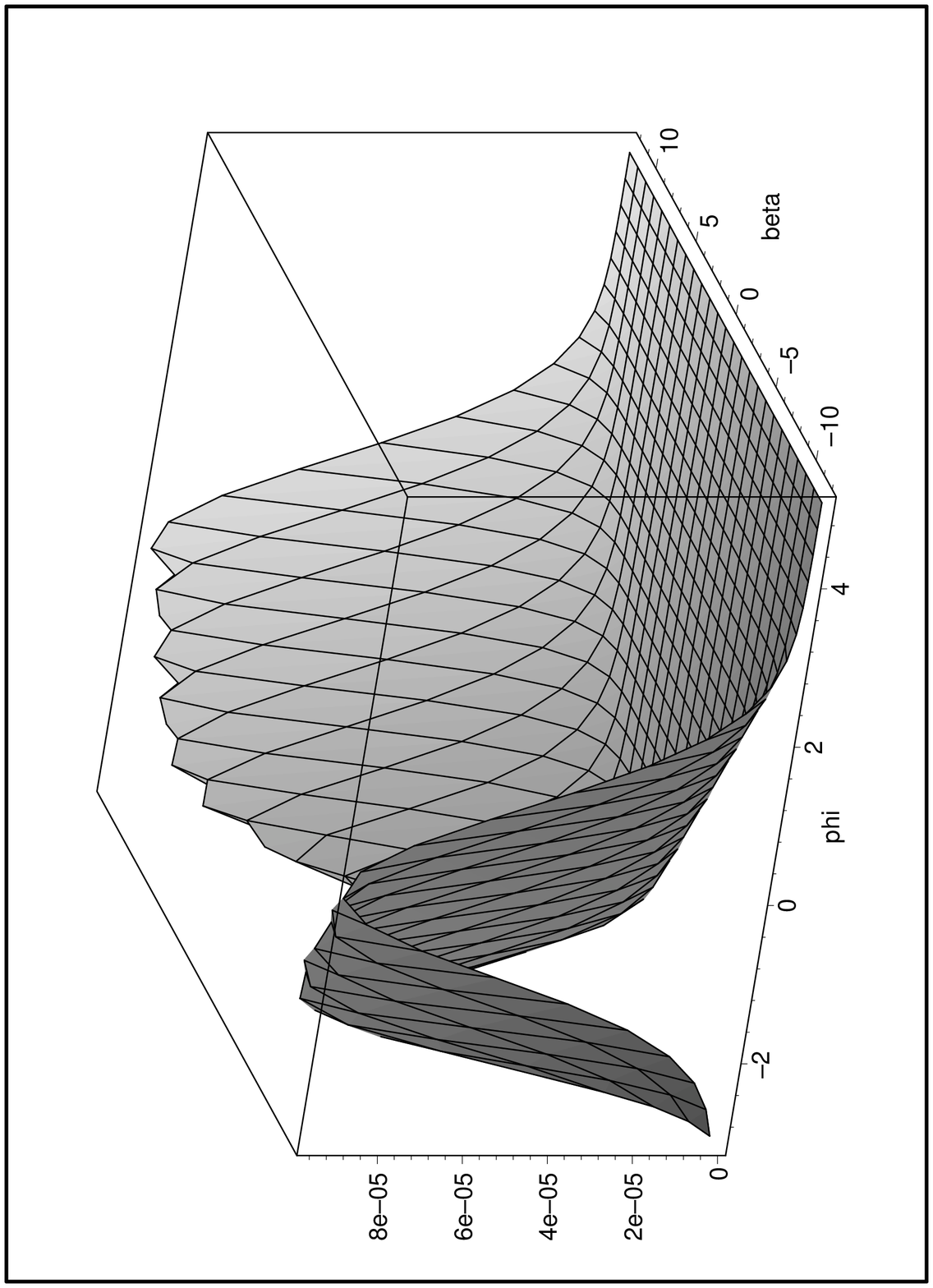}\\
\end{center}
\caption{$|\Psi(\phi, \beta)|^2$ for $\theta= 2$.} \label{wavef2}
\end{figure}
%%%%%%%%%%%%%%%%%%%%%%%%%%%%%%%%%%%%%%%%%%%%%%%%%%%%%%%%%%%%%%%%%%%%%%%

%%%%%%%%%%%%%%%%%%%%%%%%%%%%%%%%%%%%%%%%%%%%%%%%%%%%%%%%%%%%%%%%%%%%%%%

\section{Acknowledgments}

%%%%%%%%%%%%%%%%%%%%%%%%%%%%%%%%%%%%%%%%%%%%%%%%%%%%%%%%%%%%%%%%%%%%%%%
 C Mora is supported by a COFAA-IPN grant.

%%%%%%%%%%%%%%%%%%%%%%%%%%%%%%%%%%%%%%%%%%%%%%%%%%%%%%%%%%%%%%%%%%%%%%%


\begin{thebibliography}{9}

%%%%%%%%%%%%%%%%%%%%%%%%%%%%%%%%%%%%%%%%%%%%%%%%%%%%%%%%%%%%%%%%%%%%%%%

\bibitem{connes} A. Connes, M. R. Douglas and A. Schwarz, JHEP 9802 (1998) 003,
hep-th/9711162.

\bibitem{openB} C. S. Chu and P. M. Ho NUcl. Phys. B550 (1999)
151; V. Schomerus, JHEP 98026 (1999) 030, hep-th/9903205; D.
Bigatti and L. Susskind, hep-th/9908056; N. Seibert and E. Witten,
JHEP 9809 (1999) 032, hep-th/9908142 .

\bibitem{compean} H. Garc\'{\i}a-Compe\'an, O. Obreg\'on and C. Ram\'{\i}rez,
Phys. Rev. Lett. {\bf 57}, 1015 (1998).

\bibitem{cnc}  Juan M. Romero and  J. David Vergara,
 Mod.Phys.Lett. A18 (2003) 1673-1680, hep-th/0303064.

\bibitem{gnc} A. H. Chamseddine, J.Math.Phys. 44 (2003) 2534-2541, hep-th/0202137; V.
O. Rivelles Noncommutative Field Theory and Gravity,
hep-th/0212262.

\bibitem{qnc} J.M. Carmona, J.L. Cortes, J. Gamboa, F. Mendez,
JHEP 0303 (2003) 058, hep-th/0301248.

\bibitem{pm} L. O. Pimentel and C. Mora, work in preparation.

\bibitem{Gasperini} M. Gasperini, J. Maharana and G. Veneziano,
Nucl. Phys. B {\bf 472}, 349 (1996).

\bibitem{Gamboa} J. Gamboa, M. Loewe and J. C. Rojas, Phys. Rev. D62, 067901 (2001),
hetp-th/0010220.

\bibitem{Moyal} J.E. Moyal, Proc. Camb. Phil. Soc. 45 (1949) 99.

\end{thebibliography}
\end{document}